# A Unified Stochastic Particle Bhatnagar-Gross-Krook Method for Multiscale Gas Flows

Fei Fei[1], Jun Zhang[2, *], Jing Li[3], ZhaoHui Liu[3, *]

1. School of Aerospace Engineering, Huazhong University of science and technology, 430074 Wuhan, China
2. School of Aeronautic Science and Engineering, Beihang University, Beijing 100191, P.R. China.
3. State Key Laboratory of Coal Combustion, School of Energy and Power Engineering, Huazhong University of science and technology, 430074 Wuhan, China

(*corresponding authors: jun.zhang@buaa.edu.cn; zliu@hust.edu.cn)

**Abstract.** The stochastic particle method based on Bhatnagar-Gross-Krook (BGK) or ellipsoidal statistical BGK (ESBGK) model approximates the pairwise collisions in the Boltzmann equation using a relaxation process. Therefore, it is more efficient to simulate gas flows at small Knudsen numbers than the counterparts based on the original Boltzmann equation, such as the Direct Simulation Monte Carlo (DSMC) method. However, the traditional stochastic particle BGK method decouples the molecular motions and collisions in analogy to the DSMC method, and hence its transport properties deviate from physical values as the time step increases. This defect significantly affects its computational accuracy and efficiency for the simulation of multiscale flows, especially when the transport processes in the continuum regime is important. In the present paper, we propose a unified stochastic particle ESBGK (USP-ESBGK) method by combining the molecular convection and collision effects. In the continuum regime, the proposed method can be applied using large temporal-spatial discretization and approaches to the Navier-Stokes solutions accurately. Furthermore, it is capable to simulate both the small scale non-equilibrium flows and large scale continuum flows within a unified framework efficiently and accurately. The applications of USP-ESBGK method to a variety of benchmark problems, including Couette flow, thermal Couette flow, Poiseuille flow, Sod tube flow, cavity flow, and flow through a slit, demonstrated that it is a promising tool to simulate multiscale gas flows ranging from rarefied to continuum regime.

**Key words**: Stochastic Particle Method, BGK model, Multiscale Modelling, Multiscale Flows.

## 1. Introduction

Multiscale modelling of gas flows is attracting more and more attentions as a large number of gas flows encountered in modern engineering problems are inherently multiscale, especially in aerospace engineering [1, 2] and micro-electro-mechanical system (MEMS) [3]. One example is high-speed gas flows around a reentry vehicle. Assuming the characteristic length of the reentry vehicle is 1 m, the global Knudsen number (Kn, the definition is the ratio of the molecular mean free path to the characteristic length) ranges from $10^{-6}$ to $10^{-1}$ at the altitudes of 20~100 km, and correspondingly the gas flow changes from continuum to transition regime. Furthermore, if local structures such as the sharp leading edge or the microstructures on the vehicle surfaces are considered, the local Kn number spans a wider range, which will introduce a variety of thermochemical nonequilibrium phenomena and affect the flow fields around the reentry vehicle significantly. To accurately simulate such kinds of multiscale gas flows is very challenging. Although computational fluid dynamics (CFD) methods based on the Navier–Stokes (NS) equation have been successfully applied to the continuum regime, they encounter physical limitations for the simulation of gas flows far from equilibrium.

On the other hand, the direct simulation Monte Carlo (DSMC) method [4] on the molecular level is applicable to the simulation of nonequilibrium gas flows. Theoretically, DSMC is valid for the whole range of flow regime, as it can be regarded as a particle simulation method of solving Boltzmann equation. While it is popularly applied in the transition and near-continuum regime, the direct application of it to continuum regime is quite expensive due to the limitation of time steps and cell sizes. Hence, a straightforward way to construct a multiscale method is coupling DSMC method and a CFD scheme, e.g., DSMC-CFD hybrid method, where the rarefied and continuum flow regimes are solved by the DSMC and CFD methods, respectively [5-8]. However, DSMC-CFD hybrid approaches suffer from difficulties because of the amalgamation of two fundamentally different types of solvers [9]. It is very subtle to exchange information at the interface between DSMC and CFD regions.

One promising strategy for multiscale modelling is to develop a consistent solver for the whole flow regimes. Among others, one typical progress in this direction is the unified gas-kinetic scheme (UGKS) proposed by Xu and Huang [10] and discrete unified gas-kinetic scheme (DUGKS) proposed by Guo etc. [11,12], which have been successfully applied to a variety of multiscale gas flows [13-15]. For both continuum and rarefied regimes, these two methods compute the gaseous distribution functions through discrete molecular velocities. Alternatively, some researchers have made efforts to develop a particle-particle hybrid method, such as BGK-DSMC [16-19] and Fokker-Planck-DSMC [20-23] methods, where the particle simulation methods based on BGK or Fokker-Planck model are employed for the continuum regime, while DSMC method is used for the rarefied regime. It is known that BGK [24, 25] or Fokker-Planck [26] model simplifies the collision term in the Boltzmann equation, so their corresponding particle methods can achieve much higher efficiency than DSMC in the continuum regime. Compared to the UGKS and DUGKS methods, particle-particle hybrid methods are more efficient for the simulation of high speed gas flows, especially when complex physical and chemical effects are taken into account.

The stochastic particle method based on the BGK model was proposed by Macrossan [27] and Gallis and Torczynski [28] independently. Recently, the application of this method has been extended to complex gas flows [29-32]. Note that in the current stochastic particle BGK method, computational particles mimic the kinetic equations in two stages, i.e. convection and collision, which are decoupled in one calculating time step as same as that in the DSMC method. Consequently, their transport coefficients, such as viscosity and thermal conductivity, will deviate from physical values significantly if the time step is larger than the molecular mean collision time. As analyzed by Chen and Xu [33], a successful multiscale gas kinetic scheme need to inherently couple convection and collision effects when large temporal-spatial discretization is used.

Since the seminal work of Jenny etc. [34], the stochastic particle method based on the Fokker-Planck model has been developed and applied widely [35-40]. The integral solution of the Fokker-Planck model naturally couples the molecular convection and collision, and hence theoretically its viscosity and thermal conductivity can satisfy the NS solutions at large time steps [37, 40]. However, the integral solution implicitly underestimates the pressure effect when large time steps are used. To solve this problem, a macroscopic pressure term has been introduced by the authors in the multiscale temporal discretization Fokker-Planck (MTD-FP) method [40]. Although the MTD-FP method has been successfully applied to a variety of gas flows using large time steps, the combined solution of the macroscopic pressure term and the microscopic particle motions significantly affects computational efficiency and numerical stability.

In the present paper, a unified stochastic particle algorithm based on the BGK model is proposed by coupling molecular convection and collisions. Our aim is to improve the accuracy of the current stochastic particle BGK method for large temporal-spatial discretization and to develop a unified multiscale particle method in the end. Comparing with the MTD-FP method, the unified multiscale particle method presented here does not need to be solved combining with macroscopic equations.

The remainder of this paper is organized as follows. In section 2, we first review the ellipsoidal statistical Bhatnagar–Gross–Krook (ESBGK) model and the related stochastic particle method. In section 3, we present the principle and algorithm of the proposed unified stochastic particle method for multiscale gas flows. At last, several applications of the proposed method for a wide range of Kn numbers and time steps are presented in Section 4.

## 2. The stochastic particle method for ESBGK model

On the microscopic point of view, the state of gas flows is determined by the probability distribution function (PDF) $f(\mathbf{c},\mathbf{x},t)$ of gas molecules, where $\mathbf{c}$ and $\mathbf{x}$ are molecular velocity and position at time $t$, respectively. The macroscopic quantities of gas flows can be obtained from the PDF by taking averages of the corresponding microscopic quantities as follows,

$$\rho = \int f d\mathbf{c}, \qquad \rho u_i = \int c_i f d\mathbf{c},$$
$$\rho e = \frac{3}{2}p = \frac{3}{2}\rho RT = \int \frac{1}{2}C^2 f d\mathbf{c},$$
$$p_{ij} = \int C_i C_j f d\mathbf{c} = p\delta_{ij} + \sigma_{ij}, \quad \sigma_{ij} = \int C_{<i} C_{j>} f d\mathbf{c}, \qquad (1)$$
$$q_i = \frac{1}{2}\int C^2 C_i f d\mathbf{c},$$

where $\rho$ is mass density, $u_i$ is macroscopic flow velocity, $e$ is internal energy, $T$ is temperature, $R = k_B/m$ is the gas constant, $m$ is the molecular mass, $k_B$ is the Boltzmann constant. $C_i = c_i - u_i$ is the peculiar velocity of molecules. $p$ is the hydrostatic pressure, $p_{ij}$ is the pressure tensor, $\sigma_{ij}$ is the trace-free part of the pressure tensor, and $\delta_{ij}$ is the Kronecker delta function. $q_i$ is the heat flux.

In gas kinetic theory, the evolution of the PDF is governed by the kinetic equation, i.e.

$$\frac{\partial f(\mathbf{c},\mathbf{x},t)}{\partial t} + c_i \frac{\partial f(\mathbf{c},\mathbf{x},t)}{\partial x_i} = J(f), \qquad (2)$$

The left hand side of Eq. (2) refers to the change of PDF due to molecular motions in space, and external forces are omitted here for the sake of simplicity; The term $J(f)$ on the right hand side of Eq. (2) describes the change of PDF due to collisions among molecules. In the Boltzmann equation, the binary collision is assumed, and the collision term is written as

$$J_{(\text{Boltzmann})} = \int_{-\infty}^{+\infty} \int_0^{4\pi} \left( f' f_1' - f f_1 \right) g \sigma d\Omega d\mathbf{c}_1, \qquad (3)$$

where $f$ and $f_1$ are the PDF of the two colliding molecules before collision, and $f'$ and $f_1'$ are the corresponding PDF after collision. $g = |\mathbf{c} - \mathbf{c}_1|$ is the relative velocity of the colliding molecules, $\sigma$ is the differential cross-section of the binary collision, and $\Omega$ is the solid angle. As the collision term of the Boltzmann equation involves multiple integrations in velocity space, it is difficult to compute directly. To circumvent the calculation of multiple integrations, simplified models such as the BGK [24] or Fokker-Planck models [26] have been proposed to describe the binary collision using a relaxation process.

The BGK model approximates the collision term as

$$J_{(BGK)} = \upsilon(f_e - f), \tag{4}$$

where $\upsilon = p/\mu$ is the relaxation frequency, and $f_e$ is the Maxwellian distribution function,

$$f_e = \rho \left(\frac{1}{2\pi RT}\right)^{3/2} \exp\left(-\frac{C^2}{2RT}\right). \tag{5}$$

Numerical schemes including the discrete velocity and stochastic particle methods have been developed to solve the BGK model. However, the Prandtl (Pr) number determined by the original BGK model is always unity for any gas flows, and this inevitably introduces error if thermal conductivity plays an important role in gas flows. To correct the Pr number, several modified models have been developed, such as the Shakhov (SBGK) and ellipsoidal statistical BGK (ESBGK) models. The ESBGK model is proposed by Holway [41] and Cercignani [42], and it has been demonstrated to satisfy Boltzmann's H-theorem recently [25].

The ESBGK model replaces the Maxwellian distribution in Eq. (4) by a local anisotropic three-dimensional Gaussian distribution $f_G$ and uses a modified relaxation frequency $\upsilon_{ES}$ ($\upsilon_{ES} = \text{Pr} \cdot \upsilon$) as follows,

$$J_{(ESBGK)} = \upsilon_{ES}(f_G - f), \tag{6a}$$

where $f_G$ has the form as

$$f_G = \rho \left[\frac{1}{\det(2\pi\lambda_{ij})}\right]^{1/2} \exp\left(-\frac{1}{2}\lambda_{ij}^{-1} C_i C_j\right). \tag{6b}$$

And the matrix $\lambda_{ij}$ is

$$\lambda_{ij} = RT\delta_{ij} + \left(1 - \frac{1}{\text{Pr}}\right)\frac{\sigma_{ij}}{\rho}. \tag{6c}$$

Stochastic particle method for the ESBGK model (SP-ESBGK) has been developed by Gallis and Torczynski [28] and Burt and Boyd [29], respectively. Similar to the DSMC method, the molecular motions and inter-molecular collisions are decoupled into two stages in one calculating time step in the SP-ESBGK method, and the corresponding governing equations for these two stages can be written as convection: $\left[\dfrac{Df}{Dt}\right]_{convection} = 0$, (7a) relaxation: $\left[\dfrac{\partial f}{\partial t}\right]_{relaxation} = \upsilon_{ES}\left(f_G - f\right)$. (7b) where $D/Dt = \partial/\partial t + c_i \partial/\partial x_i$. According to Eq. (7), the distribution function $f$ is updated through particle convection and relaxation process, i.e.

particle convection: $f^*(\mathbf{c}, \mathbf{x}_{n+1}, t_{n+1}) = f(\mathbf{c}, \mathbf{x}_n, t_n)$, (8a)

relaxation procedure: $f^{n+1} = f^* e^{-\upsilon_{ES}\Delta t} + \left(1 - e^{-\upsilon_{ES}\Delta t}\right) f_G^*$, (8b)

where $t_{n+1} = t_n + \Delta t$, $\mathbf{x}_{n+1} = \mathbf{x}_n + \mathbf{c}\Delta t$, and $\Delta t$ is the time step. For short, the brackets as well as the contents in them have been omitted in Eq. (8b). In the following, the quantities at $t_n$, $t_{n+1}$ and the time just after particle convection stage are denoted with the superscript n, n+1 and an asterisk, respectively.

The main difference between the DSMC and SP-ESBGK methods is the treatment of the collision process as shown in Eq. (8b). In the SP-ESBGK method, the number of particles selected ($N_s$) for collisions in each cell depends on the relaxation frequency and the time step,

$$N_s = \text{int}\left\{N_c\left[1 - \exp(-\upsilon_{ES}\Delta t)\right]\right\},$$ (9)

where $N_c$ is the number of particles in a computational cell, and the operator "int" returns the nearest integer. The selected particles are assigned new thermal velocities from a Maxwellian distribution by

$$C_i^* = \cos\left(2\pi R_{f1}\right)\sqrt{-\ln\left(R_{f2}\right)} \cdot \sqrt{2k_B T / m},$$ (10)

where $R_{f1}$ and $R_{f2}$ are independent random numbers between 0 and 1. The velocities of particles that have not been preselected remain unchanged. According to ESBGK model, the assigned thermal velocities should be modified to conform to the Gaussian distribution $f_G$. Gallis and Torczynski [28] proposed that the modified velocities $C_i$ can be determined from the resampled velocities $C_i^*$ as

$$C_i = S_{ij} \cdot C_j^*,$$ (11a)

where $S_{ij}$ is given by

$$S_{ij} = \delta_{ij} - \dfrac{1-\text{Pr}}{2\text{Pr}}\left[\dfrac{m}{k_B T}\dfrac{p_{ij}}{\rho} - \delta_{ij}\right].$$ (11b)

For the sake of clarity, the numerical implementation of the SP-ESBGK method is briefly summarized in Table 1.

Table 1. Outline of the algorithm of the SP-ESBGK method

| | |
|---|---|
| 1. | Advect the particles (similar to DSMC). |
| 2. | Apply boundary conditions (similar to DSMC). |
| 3. | Assign new thermal velocities to selected particles (Eqs. (9) and (10)). |
| 4. | Modify the velocities to conform to the Gaussian distribution $f_G$ (Eq. (11)). |
| 5. | Sample the results (similar to DSMC). |

## 3. Unified stochastic particle method for ESBGK model

### 3.1 The governing equations

For large scale gas flows, the numerical viscosity and thermal conductivity of the SP-ESBGK method increase with time steps in analogy to the DSMC method. The reason for this is that molecular motions and inter-molecular collisions are implemented separately. As analyzed by Chen and Xu [33], in order to recover the NS solution in the continuum limit at large time steps, both the effects of molecular motions and inter-molecular collisions need to be considered in the convection and relaxation procedures. In this paper, we proposed a unified stochastic particle method based on ESBGK model (USP-ESBGK) for the simulation of multiscale gas flows, and the governing equations are assumed as follows,

convection:
$$\left[\frac{Df}{Dt}\right]_{convection} = J, \quad (12a)$$

relaxation:
$$\left[\frac{\partial f}{\partial t}\right]_{relaxation} + \left[\frac{Df}{Dt}\right]_{convection} = \upsilon_{ES}(f_G - f). \quad (12b)$$

Comparing with the governing equations of SP-ESBGK method (Eq. (7)), it can be seen that a collision term $J$ and the convection term $[Df/Dt]_{convection}$ are added to the right hand side in Eq. (12a) and the left hand side in Eq. (12b), respectively. If $J$ in Eq. (12a) is taken to be the exact collision term of the ESBGK model as Eq. (6a), the convection stage (Eq. (12a)) is identical to the ESBGK equation, and the relaxation stage (Eq. (12b)) reads

$$\left[\frac{\partial f}{\partial t}\right]_{relaxation} = 0. \quad (13)$$

Consequently, the exact PDF of the gas flows can be obtained directly from the convection stage combined with collision effect in theory. However, it is difficult to realize in the stochastic particle methods, because the PDF as well as the collision term of the ESBGK model cannot be calculated with an explicit formulation using simulated molecules.

Therefore, an approximated collision term of $J$ is assumed and implemented in the USP-ESBGK method,

$$J = J_{(USP-ESBGK)} = \upsilon P_{ne} \left( f_e - f_{|Grad} \right), \tag{14}$$

where $P_{ne}$ is a parameter corresponding to the degree of rarefication, and the 13 moments Grad's distribution function $f_{|Grad}$ is applied to close the collision term with the form as

$$f_{|Grad} = f_{|13} = f_e \left[ 1 + \frac{\sigma_{ik}}{2p} \frac{C_{<i}C_{k>}}{\theta} + \frac{2}{5} \frac{q_k}{p\theta} \Pr C_k \left( \frac{C^2}{2\theta} - \frac{5}{2} \right) \right], \tag{15}$$

where $\theta = k_B T/m$, and the Pr number in the last term is used to correct the thermal conductivity in the BGK model.

Specifically, the parameter $P_{ne}$ in Eq. (14) is

$$P_{ne} = e^{-Kn_{GLL,MAX}/Kn_c}, \tag{16}$$

where $Kn_c$ is a reference Knudsen number (selected as 0.1 in the present paper), and $Kn_{GLL,MAX}$ is the maximum value of the gradient-length local (GLL) Knudsen number suggested by Wang and Boyd [43] to evaluate the degree of non-equilibrium effect, i.e.,

$$Kn_{GLL,MAX} = \max \left( Kn_{GLL,\rho}, Kn_{GLL,T}, Kn_{GLL,u} \right). \tag{17}$$

The GLL Knudsen number in the above equation is defined as

$$Kn_{GLL,Q} = \frac{\lambda}{Q} \left| \frac{dQ}{dl} \right|, \tag{18}$$

where Q could be any flow property such as density, temperature, or flow velocity.

Using the assumed collision term as Eq. (14), the governing equations for the USP-ESBGK method (Eq. (12)) can be rewritten as,

convection: $\qquad \left[ \dfrac{Df}{Dt} \right]_{convection} = J_{(USP-ESBGK)}, \tag{19a}$

relaxation: $\qquad \left[ \dfrac{\partial f}{\partial t} \right]_{collision} = \upsilon_{ES} \left( f_G - f \right) - J_{(USP-ESBGK)}. \tag{19b}$

The term on the right hand side of Eq. (19a) represents the collision effect near equilibrium. This effect is combined and calculated in the convection stage, and it makes the corresponding results converge to the NS limit in large temporal-spatial scales (see Appendix A). On the other hand, the term on the right hand side of Eq. (19b) represents the collision effect far from equilibrium. This effect is solved in the relaxation stage similar to the SP-ESBGK method, and it captures the non-equilibrium solutions in small temporal-spatial scales (see Appendix B). The ratio of these two collision effects is adaptively determined by the parameter $P_{ne}$ as a function of the local Knudsen number.

### 3.2 The numerical method using stochastic particle

Based on the governing equations (19), the USP-ESBGK method is implemented using stochastic particles. Similar to the SP-ESBGK method, it contains two main stages, i.e. particle convection [Eq. (19a)] and collision relaxation processes [Eq. (19b)]. The numerical implements of these two stages are described as follows.

### 3.2.1 Calculation of particle convection

Similar to the DSMC method, each computational particle is initially assigned a position $\mathbf{x}_0$ and a velocity $\mathbf{c}$ according to the initial conditions of the flow fields. Additionally, a particle weight $W$ is assigned to each computational particle, and it is equal to 1.0 for the equilibrium PDF.

After initialization, the particle convection of Eq. (19a) can be numerically solved by applying the trapezoidal rule for the collision term as

$$f^*(\mathbf{c}, \mathbf{x}_{n+1}, t_{n+1}) - f(\mathbf{c}, \mathbf{x}_n, t_n) = \frac{\Delta t}{2} \left[ J^*_{(USP-ESBGK)}(\mathbf{c}, \mathbf{x}_{n+1}, t_{n+1}) + J_{(USP-ESBGK)}(\mathbf{c}, \mathbf{x}_n, t_n) \right]. \tag{20}$$

Introducing auxiliary PDF $\tilde{f}$ and $\hat{f}$ as same as that in the DUGKS method [11], i.e.

$$\tilde{f}^* = f^* - \frac{\Delta t}{2} J^*_{(USP-ESBGK)}, \tag{21}$$

and

$$\hat{f}^n = f^n + \frac{\Delta t}{2} J^n_{(USP-ESBGK)}. \tag{22}$$

where superscript n and asterisk represent to the quantities at $t_n$ and after particle convection stage, respectively. Substituting Eqs. (21) and (22) into Eq. (20) yields

$$\tilde{f}^*(\mathbf{c}, \mathbf{x}_{n+1}, t_{n+1}) = \hat{f}(\mathbf{c}, \mathbf{x}_n, t_n). \tag{23}$$

It means that if the PDF at time $t_n$ ($\hat{f}^n$) has been reconstructed from Eq. (22), the auxiliary PDF of computational particles after convection at time $t_{n+1}$ ($\tilde{f}^*$) is determined by Eq. (23), and then the real PDF of computational particles after convection ($f^*$) can be obtained as

$$f^* = \tilde{f}^* + \frac{\Delta t}{2} J^*_{(USP-ESBGK)}. \tag{24}$$

Therefore, the key point in the particle convection stage is to construct $\hat{f}^n$ and $f^*$ based on the known distribution functions $f^n$ and $\tilde{f}^*$, respectively (see Eqs. (22) and (24)). In the following, we first illustrate how to construct $\hat{f}^n$ from $f^n$ according to Eq. (22). The basic idea is to add extra computational particles, whose distribution function is required to satisfy $\Delta t\, J^n_{(USP-ESBGK)}/2$. In the present scheme, the number of added particles $N_a$ in a computational cell is chosen as

$$N_a = \begin{cases} (\rho^n V_{cell})/(mr_p N_p) & \Delta t \cdot \upsilon_{ES} \geq 1 \\ \Delta t \upsilon_{ES}(\rho^n V_{cell})/(mr_p N_p) & \Delta t \cdot \upsilon_{ES} < 1 \end{cases}, \quad (25)$$

where $\rho^n$ is the gas density at time $t_n$, $V_{cell}$ is the cell volume, $N_p$ is the number of real molecules represented by one computational particle of weight 1.0, and $r_p$ is a scale coefficient to control the number of added particles. Considering computational efficiency, $r_p$ is usually chosen larger than 1.0, and here we chose $r_p = 10.0$.

The velocities of these added particles are sampled from Maxwell distribution $f_e^n/\rho$. Additionally, in order to ensure the PDF of these added particles to satisfy $\Delta t\, J^n_{(USP-ESBGK)}/2$, the particle weight of each added particle is chosen as

$$W^n = \frac{r_p\left(\Delta t\, J^n_{(USP-ESBGK)}/2\right)}{f_e^n} = -\frac{r_p \upsilon P_{ne} \Delta t}{2}\left[\frac{\sigma^n_{ij}}{2p^n}\frac{C_{<i}C_{j>}}{\theta^n} + \frac{2}{5}\frac{q^n_i}{p^n\theta^n}\Pr C_i\left(\frac{C^2}{2\theta^n}-\frac{5}{2}\right)\right]. \quad (26)$$

The macro valuables in above equations, such as $\rho$, $\theta$, $\sigma_{ij}$ and $q_i$, are obtained following Eq. (1). The procedure of constructing $f^*$ is similar to the construct of $\hat{f}^n$ as shown in Eqs. (25) and (26), except that the superscript n needs to be replaced by an asterisk. The sampling of macro valuables in constructing $f^*$ will be discussed in detail in section 3.2.3.

### 3.2.2 Calculation of collision relaxation process

Once $f^*$ is obtained, the distribution function $f^{n+1}$ is updated by a temporal integrating of Eq. (19b),

$$f^{n+1} = f^* e^{-\upsilon_{ES}\Delta t} + \left(1-e^{-\upsilon_{ES}\Delta t}\right)\int_{t_n}^{t_{n+1}} \frac{\upsilon_{ES} e^{\upsilon_{ES}(t-t_n)}}{\left(e^{\upsilon_{ES}\Delta t}-1\right)} f_G(t)dt - \left(1-e^{-\upsilon_{ES}\Delta t}\right)/\upsilon_{ES} \cdot J^*_{(USP-ESBGK)}. \quad (27)$$

The first and second terms on the right hand side of Eq. (27) is computed similar to that in the SP-ESBGK method (see Eq. (8b)).

First, $N_s$ particles are selected from the computational cell (see Eq. (9)), while the remaining particles are unchanged. However, since computational particles have different weights in the USP-ESBGK method, only a fraction of particles ($N_s^{(USP-ESBGK)}$) are chosen to assign a new particle weight as 1.0. The number $N_s^{(USP-ESBGK)}$ is determined as

$$N_s^{(USP-ESBGK)} = \text{int}\left(\sum_{k=1}^{N_s} W_k\right) \tag{28}$$

The other $\left(N_s - N_s^{(USP-ESBGK)}\right)$ particles are deleted from the computational cell. Note that the particle deletion in this step exactly offsets the adding of computational particles in the convection stage. Therefore, the total number of computational particles keeps constant throughout the whole simulation.

Second, the velocities of reassigned particles $N_s^{(USP-ESBGK)}$ are calculated from Gauss distribution $f_G$, which is a function of $\rho$, $T$ and $\sigma_{ij}$. It should be noted that the determination of $f_G(t)$ and the integral formulation of the second term of Eq. (27) in the USP-ESBGK method is different from that in the SP-ESBGK method. In the SP-ESBGK method, $\rho$, $T$ and $\sigma_{ij}$ as well as $f_G$ is assumed to be constant due to small calculating time steps, and hence the second term on the right hand side of Eq. (27) can be integrated and sampled directly (see Eq. (8b)). However, in the USP-ESBGK method, the time steps have a wide range (could be much larger than molecular collision time), and thus $\sigma_{ij}$ and $f_G$ cannot be assumed constant in one calculating time step. Consequently, the time integration in Eq. (27) needs to be solved numerically, and here Monte Carlo method is employed to get the solution.

To this end, we define a distribution function $g_{time}(t)$ with the form of

$$g_{time}(t) = \frac{\upsilon_{ES} e^{\upsilon_{ES}(t-t_n)}}{\left(e^{\upsilon_{ES}\Delta t} - 1\right)}, \tag{29}$$

where $\int_{t_n}^{t_{n+1}} g_{time}(t)dt = 1$. A certain time instant $t$ in the range of $t_n$ to $t_{n+1}$ is sampled first using Monte Carlo method,

$$t = \ln\left[R_{ft}\left(e^{\upsilon_{ES}\Delta t} - 1\right) + 1\right]/\upsilon_{ES}, \tag{30}$$

where $R_{ft}$ is random number between 0 and 1. After the time instant $t$ is known, $f_G(t)$ can be determined from the values of $\rho(t)$, $T(t)$ and $\sigma_{ij}(t)$. Note that $\rho(t)$ and $T(t)$ are constant during the relaxation process, while $\sigma_{ij}(t)$ is obtained by taking moments of Eq. (19b), i.e.,

$$\left[\frac{\partial \sigma_{ij}}{\partial t}\right]_{collision} = -\upsilon \sigma_{ij} + \upsilon P_{ne} \sigma_{ij}^*. \tag{31}$$

And hence $\sigma_{ij}(t)$ is determined by the time integration of above equation, i.e.,

$$\sigma_{ij}(t) = e^{-\upsilon t}\sigma_{ij}^* + P_{ne}\left(1-e^{-\upsilon t}\right)\sigma_{ij}^* = \left[P_{ne} + \left(1-P_{ne}\right)e^{-\upsilon t}\right]\sigma_{ij}^*. \tag{32}$$

where $\sigma_{ij}^*$ is sampled from computational particles (see Eq. (33) in next section).

Finally, the last term in Eq. (27) is constructed by adding particles in analogy to Eq. (22) in the convection stage, except that the PDF of added particles is required to satisfy $-\left(1-e^{-\upsilon_{ES}\Delta t}\right)/\upsilon_{ES} \cdot J_{(USP-ESBGK)}^*$ instead of $\Delta t\, J_{(USP-ESBGK)}^n/2$.

**3.2.3 Implementations of the USP-ESBGK method and some technique details**

Similar to the SP-ESBGK method, the implementations of the USP-ESBGK method are summarized in Table 2.

Table 2. Outline of the algorithm of the USP-ESBGK method

| | |
|---|---|
| 1. | Assign initial particles in the computational domain (similar to DSMC). |
| 2. | Arrange additional particles for $\hat{f}^n$, $N_a$ particles are added with weights $W^n$ (Eq. (26)). |
| 3. | Advect the particles and apply boundary conditions (similar to DSMC). $\tilde{f}^*$ is evaluated. |
| 4. | Arrange additional particle for $f^*$, $N_a$ particles are added with weights $W^*$ (Eq. (26)). |
| 5. | $N_s$ particles are selected (Eq.(9)), and $N_s^{(USP-ESBGK)}$ particles are reassigned. |
| 6. | Sampling the velocities of reassigned particles to conform to $f_G(t)$ (similar to SP-ESBGK). |
| 7. | Arrange additional particle to update $f^{n+1}$, $N_a$ particles are added with weights $W' = -W^* \cdot 2\left(1-e^{-\upsilon_{ES}\Delta t}\right)/\left(\upsilon_{ES}\cdot \Delta t\right)$ (Eq. (27)). |
| 8. | Sample the result (similar to DSMC). |

Note that after particle motions in step 3, the distribution function of computational particles satisfies $\tilde{f}^*$, while the sampling of additional particles in step 4 is based on $f^*$. Therefore, the macro quantities required in the construction of PDF in step 4 cannot directly calculated from Eq. (1). Instead, they are computed from $\tilde{f}^*$ as

$$\rho^* = \int \tilde{f}^* d\mathbf{c}, \qquad \rho^* u_i^* = \int c_i \tilde{f}^* d\mathbf{c},$$

$$\rho^* e^* = \frac{3}{2} p^* = \frac{3}{2} \rho^* RT^* = \int \frac{1}{2} C^2 \tilde{f}^* d\mathbf{c},$$

$$\sigma_{ij}^* = \left(\int C_{<i} C_{j>} \tilde{f}^* d\mathbf{c}\right) \Big/ \left(1 + \frac{\Delta t}{2} \upsilon P_{ne}\right), \qquad (33)$$

$$q_i^* = \left(\frac{1}{2} \int C^2 C_i \tilde{f}^* d\mathbf{c}\right) \Big/ \left(1 + \frac{\Delta t}{2} \upsilon \Pr P_{ne}\right).$$

For steady flows, an exponentially weighted time averaging method [34] is used to reduce statistical noise in sampling. Specifically, the macro variables $Q$ is calculated as

$$Q(t) = \frac{n_a - 1}{n_a} Q(t - \Delta t) + \frac{1}{n_a} \frac{N_p}{V} \sum_{k=1}^{N_c} s_k(t), \qquad (34)$$

where $n_a$ is the time steps used for averaging, and $s_k$ is the corresponding microscopic variables.

## 4. Numerical cases

In this section, four 1-D and two 2-D benchmark problems, including Couette flow, thermal Couette flow, Poiseuille flow, Sod tube flow, cavity flow, and flow through a slit, are investigated using the proposed USP-ESBGK model. In all of these cases, the flow medium is Argon gas, whose viscosity depends on temperature with a power law of the form

$$\mu = \mu_{ref} \left(T/T_{ref}\right)^{\omega}, \qquad (35)$$

where $T_{ref}$ is the reference temperature, $\mu_{ref}$ is the reference viscosity, and $\omega$ is the viscosity exponent. For steady flows, exponentially weighted moving time averaging is used, and we select $n_a = 1000$ in Eq. (34). About 500 computational particles are arranged in each cell initially. Additionally, two critical parameters, i.e. the mean free path and the mean collision time, are calculated as

$$\lambda = \frac{16}{5\sqrt{2\pi}} \frac{\mu}{\sqrt{\rho p}}, \qquad (36)$$

$$\tau_c = \lambda \Big/ \sqrt{\frac{8 k_B T}{\pi m}}. \qquad (37)$$

### 4.1 Couette and thermal Couette flows

The Couette flow is a steady flow driven by two infinite and parallel plates moving oppositely along their planes. In our simulations, the Argon gas molecules is initially set up at the standard condition (p=1 atm, and T=273 K), and the plates move oppositely at the speed of $U_{wall} = 20 m/s$. The distance between the plates is H, and $Kn = \lambda / H = 0.01$. The upper and lower plates keep the temperature of $T_{wall} = 273K$, and fully diffusive boundary condition was employed. The viscosity exponent $\omega$ is 0.81. For a direct comparison, the shear stress $\Gamma_{xy}$ is calculated by DSMC, SP-ESBGK and USP-ESBGK methods, respectively. The time steps of these methods vary from $\Delta t / \tau_c = 0.2$ to 10.0. The number of uniform computational cells is 100 in the USP-ESBGK method, and 200 in the DSMC and SP-ESBGK methods for all cases. As shown in Figure 1(a), the shear stress of the USP-BGK method is almost independent of the chosen time steps. However, the results of shear stress predicted by the DSMC and SP-ESBGK methods increase with time steps significantly, due to decoupled molecular motions and collisions in one time step.

The thermal Couette flow is driven by a temperature difference between two parallel plates. The bottom and top plates have temperatures $T_{wall} + \Delta T$ and $T_{wall} - \Delta T$, respectively, where $\Delta T = 10K$. The other computational parameters are the same as those in the Couette flows. The results of heat flux obtained by these three methods are compared in Figure 1(b). Similarly, the result of USP-ESBGK method is independent of the chosen time steps, while the results of DSMC and SP-ESBGK become larger than the real values when the time step is larger than molecular collision time. These results indicate that the coupled molecular convection and collision in the USP-ESBGK method improves the ability of getting accurate transport properties, even in large time steps.

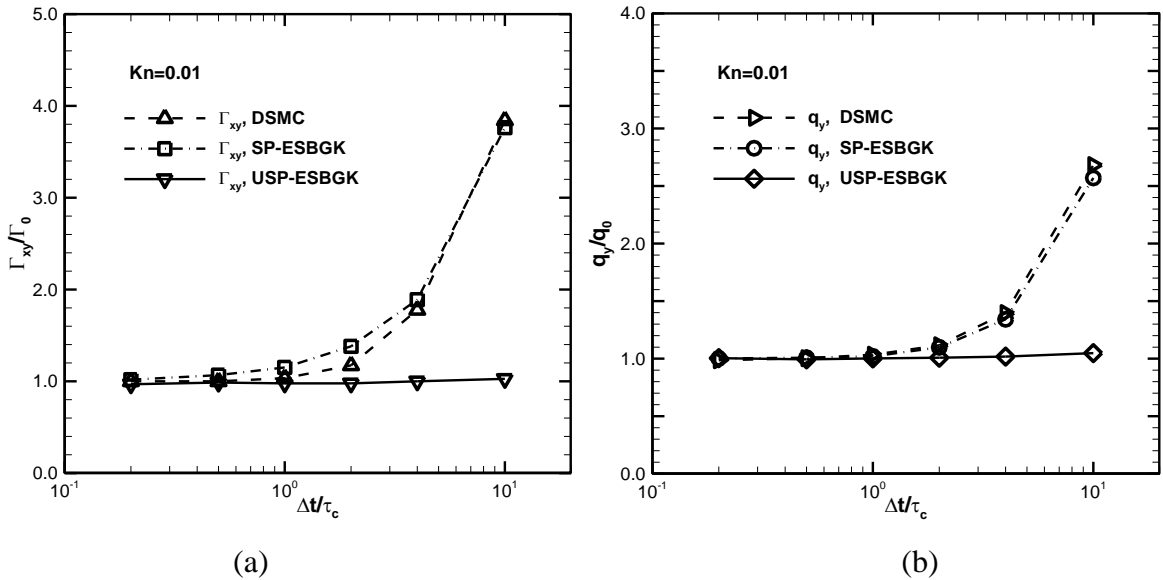

(a)          (b)

**Figure 1**: The comparison of the shear stress for the Couette flow (a) and heat flux for the thermal Couette flow (b), respectively. The Kn number is 0.01.

## 4.2 Poiseuille flow

The Poiseuille flow is confined between two infinite and parallel plates and is driven by a pressure gradient $dp/dx$ along the plates. The temperature of the upper and lower plates is fixed at 273 K, and fully diffusive boundary condition is employed for these two plates. Similar to the Couette flows, the Argon gas is initially set up at the standard condition (p=1 atm and T=273 K). The viscosity exponent $\omega$ is 0.81. The other computational parameters are shown in Table 3, where $N_{cell}$ is the number of uniform computational cell.

**Table 3:** Computational parameters of the Poiseuille flows

| Cases | Kn | $dp/dx\,(Pa\,m^{-1})$ | N$_{cell}$ (USP-ESBGK/DSMC) | $\Delta t / \tau_c$ (USP-ESBGK/DSMC) |
|---|---|---|---|---|
| 1 | 0.100 | $2.70 \times 10^{10}$ | 50/50 | 0.2/0.2 |
| 2 | 0.020 | $1.60 \times 10^{9}$ | 70/250 | 0.5/0.2 |
| 3 | 0.004 | $6.68 \times 10^{7}$ | 100/1000 | 2.0/0.2 |
| 4 | 0.001 | $4.00 \times 10^{6}$ | 100/NS solution | 5.0/NS solution |

In the continuum regime, if no-slip boundary condition is applied, the NS solutions for the velocity and temperature distributions along the direction normal to the plates are as follows [45],

$$U = \frac{dp}{dx}\frac{Hy-y^2}{2\mu}, \tag{38}$$

$$T = T_{wall} + \frac{1}{12\mu\kappa}\frac{dp}{dx}^2\left[\frac{H^4}{16}-\left(y-\frac{H}{2}\right)^4\right]. \tag{39}$$

In Figure 2, the velocity and temperature profiles obtained by the USP-ESBGK method are shown. It can be seen from Figs. 2(a-f) that the results obtained by the USP-ESBGK method agree well with the corresponding DSMC results for the cases 1-3, except that there is small deviation for the temperature distribution at Kn=0.1 as shown in Fig. 2(b), due to the limitation of the simplified collision term in the ESBGK model [46]. Note that an accurate DSMC calculations require that the cell sizes are smaller than molecular mean free path and the time steps are smaller than mean collision time, while the USP-ESBGK method can get accurate results using much larger cell sizes and time steps. For the case 4 with Kn=0.001, we compare the result of USP-ESBGK method with the NS solutions as it is in the continuum regime, and they are also consistent as shown in Figs. 2(g-h).

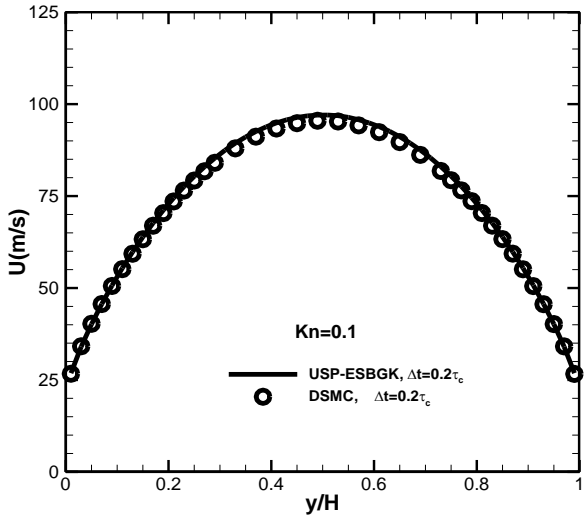

(a)

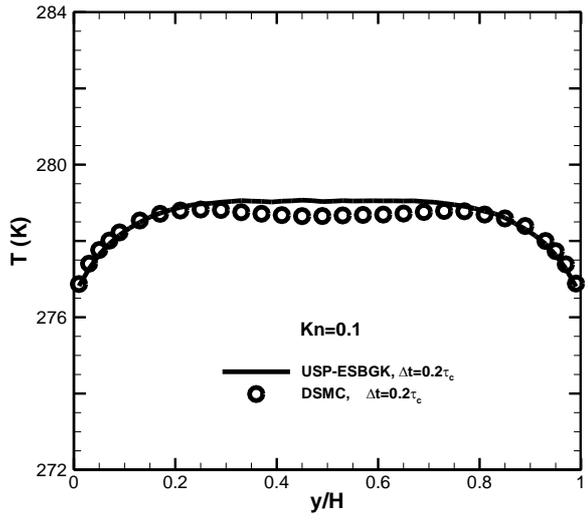

(b)

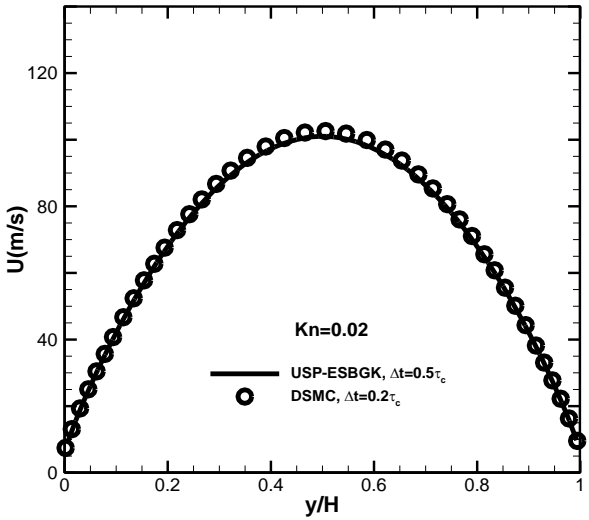

(c)

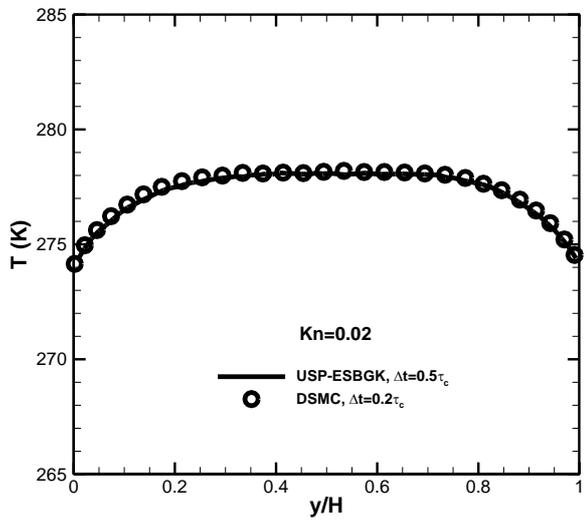

(d)

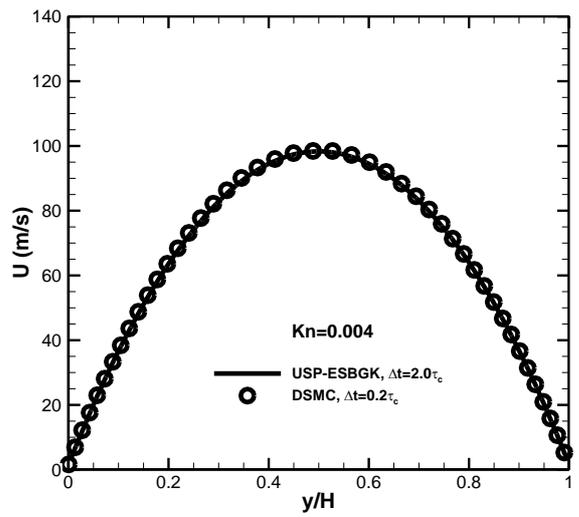

(e)

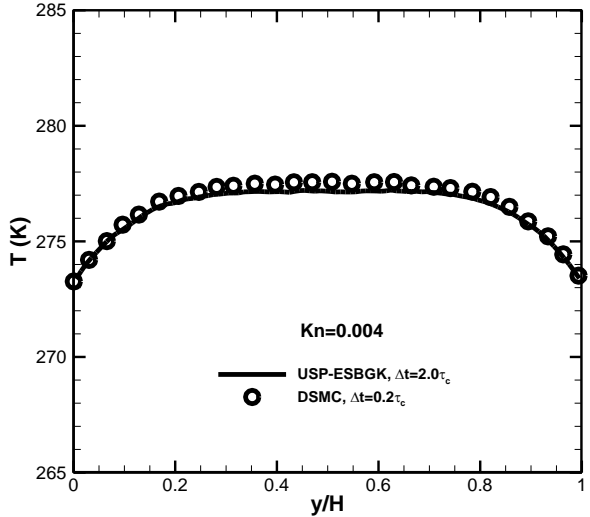

(f)

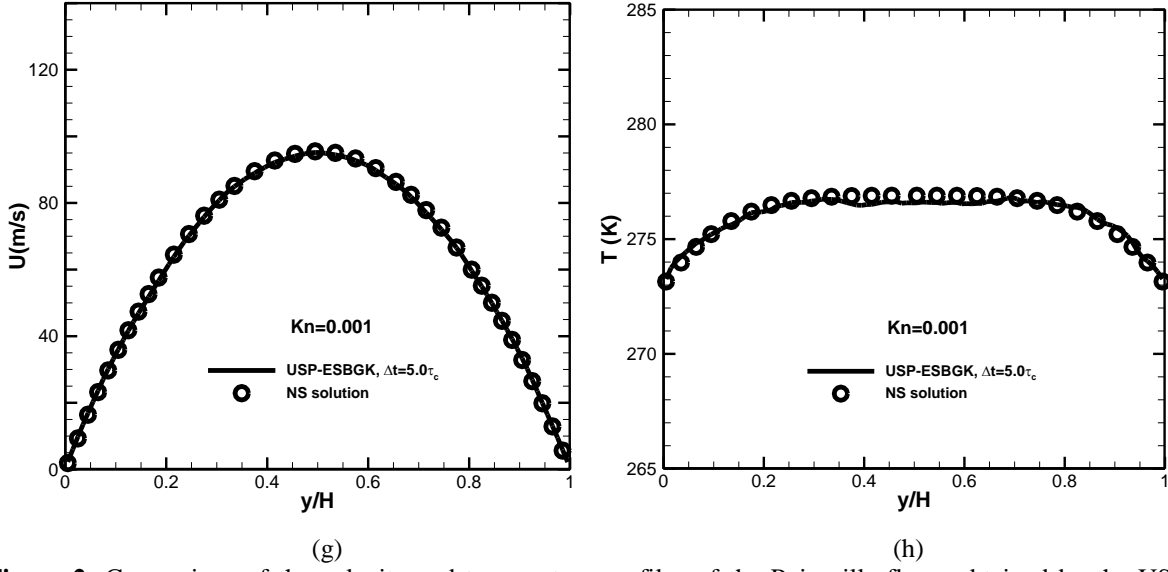

(g)                                              (h)

**Figure 2:** Comparison of the velocity and temperature profiles of the Poiseuille flows obtained by the USP-ESBGK method, the DSMC method, and the NS solutions.

To investigate the effect of time step on the results obtained by the USP-ESBGK and DSMC method, we further calculate the case 3 using different time steps from $0.2\tau_c$ and $5.0\tau_c$, and keep the cell numbers the same as that shown in Table 3. It can be seen from Fig. 3 that DSMC method underestimates the maximum velocity when $\Delta t / \tau_c > 0.5$. The reason for this is that as the time step increases, the viscosity predicted by DSMC method become larger than the physical values. In contrast, the viscosity predicted by the USP-ESBGK method are almost independent of the time steps. Therefore, USP-ESBGK method can be applied using a wider range of time steps, as shown in Figure 3.

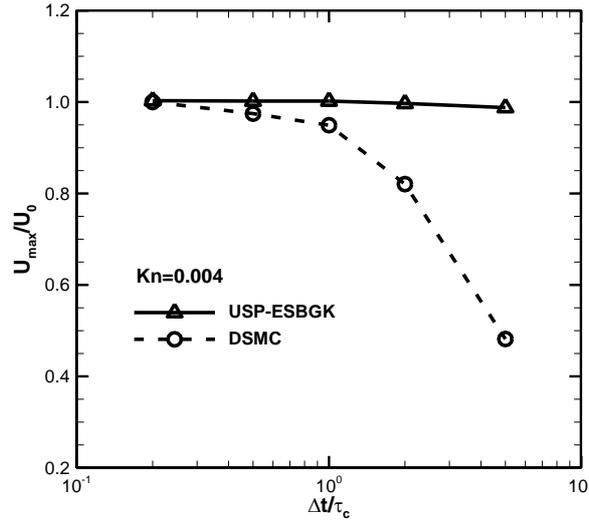

**Figure 3:** Comparison of the normalized maximum velocity of the Poiseuille flow obtained by USP-ESBGK and DSMC method with different time steps. The Kn number is 0.004.

### 4.3 Sod tube flow

The Sod's 1D shock tube problem [47] is a typical multiscale gas flow. Two cases from ref. [48] are selected and calculated using USP-ESBGK method. The length of the tube is 1 m for both cases, and the boundary conditions at the left and right ends of the tube are open boundaries. At x=0.5m, there is an initial discontinuity of density, and the initial macro velocities are zero at the whole computational domain. The initial temperatures are $T_l = 273.008012K$ and $T_r = 273.00641K$ in each chamber, and the subscripts "l" and "r" denote the left and right chambers, respectively. The other computational parameters are given in Table 4. Note that three different time steps and cell sizes are used for each case, and $\tau_{c,l}$ represents the initial mean collision time for the left chamber. To investigate the unsteady process, both cases are simulated up to the time $t_{final} = 6.8 \times 10^{-4} s$. Different from steady flows, a large number of particles are employed here to reduce statistical noise. Similar to the Fokker-Planck-DSMC method employed in ref. [20], $2.5 \times 10^4$ and $20 \times 10^4$ computational particles are initially arranged in one computational cell in the right and left chambers, respectively. As shown in Figs. 4 and 5, the USP-ESBGK results of velocity, temperature, and density with all time steps agree well with DSMC results given by ref. [48].

**Table 4:** Computational parameters of the Sod tube flows

| Case | $\rho_l (kg\,m^{-3})$ | $\rho_r (kg\,m^{-3})$ | $N_{cell}$ | $\Delta t / \tau_{c,l}$ |
|---|---|---|---|---|
| 1 | $10^{-5}$ | $0.125 \times 10^{-5}$ | 400 | 0.2 |
|   |           |                        | 100 | 0.8 |
|   |           |                        | 50  | 1.5 |
| 2 | $10^{-4}$ | $0.125 \times 10^{-4}$ | 1200 | 0.2 |
|   |           |                        | 240  | 1.0 |
|   |           |                        | 60   | 4.0 |

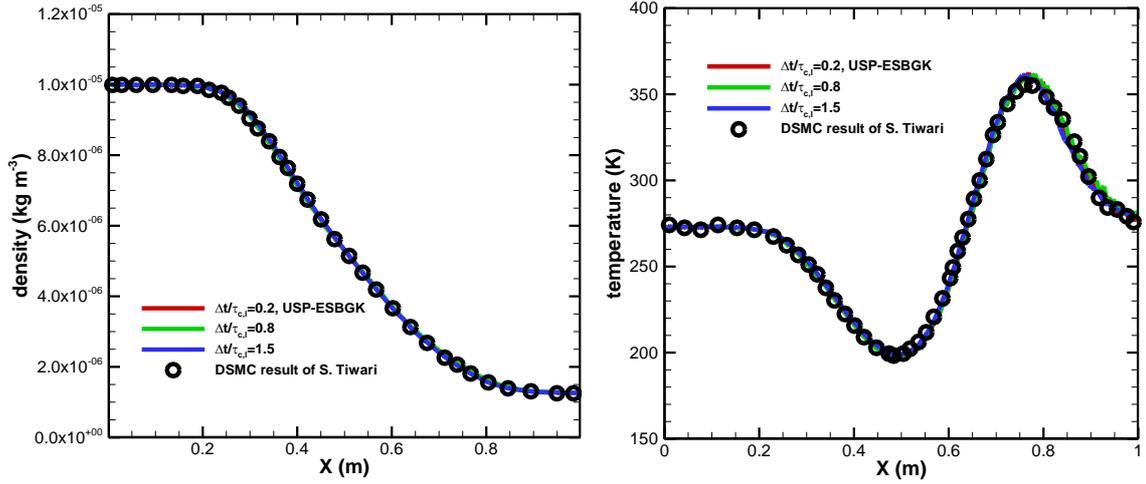

(a)                                 (b)

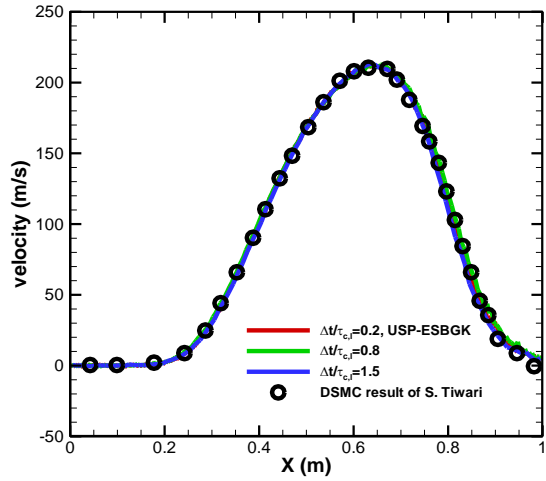

(c)

**Figure 4:** (Color online) Sod tube case 1. (a) density; (b) temperature; and (c) velocity at the final time $t_{final} = 6.8 \times 10^{-4} s$. The lines are USP-ESBGK results for three different time steps, and the symbols refer to the data of the DSMC method by S. Tiwari [48].

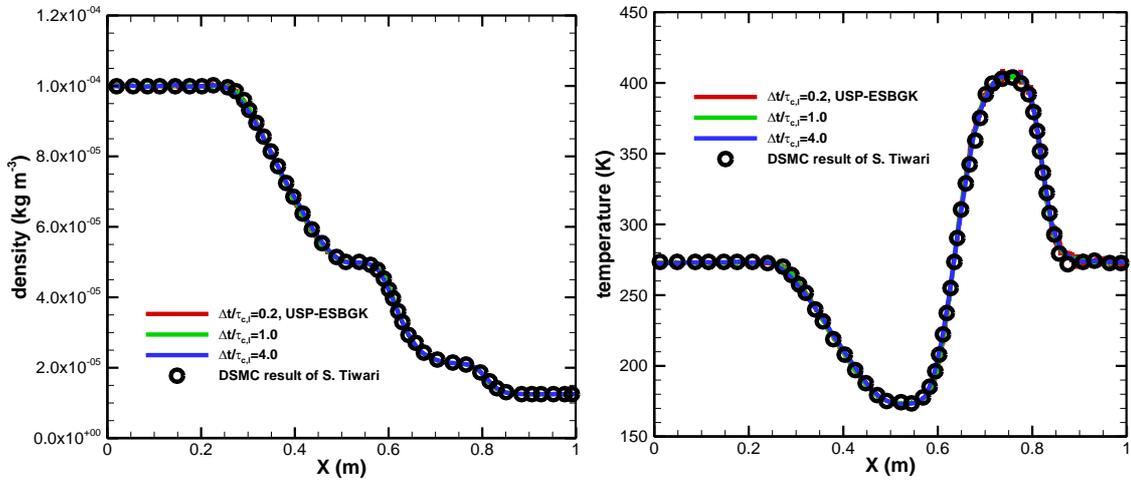

(a)                                 (b)

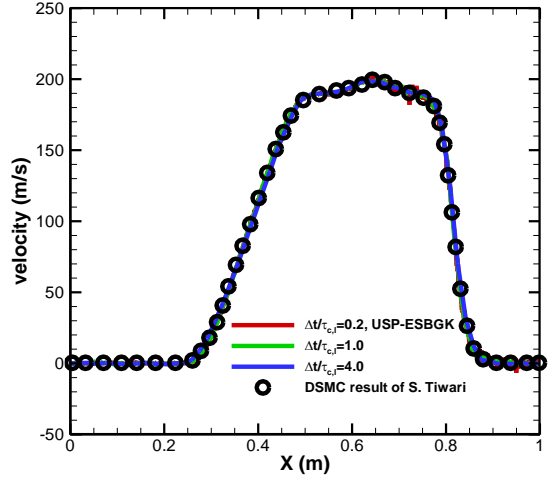

(c)

**Figure 5:** (Color online) Sod tube case 2. (a) density; (b) temperature; and (c) velocity at the final time $t_{final} = 6.8 \times 10^{-4} s$. The lines are USP-ESBGK results for three different time steps, and the symbols refer to the data of the DSMC method by S. Tiwari [48].

### 4.4 Square Cavity flow

Square cavity flow is a flow driven by the lid side moving at speed of $U_{lid}$ along the plate direction (see Fig. 6), while the other three sides keep stationary. The four sides of the cavity are all diffusively reflective and have the same temperature as the initial gas, $T_{wall} = T_0 = 273K$. The flow medium is argon gas at the standard condition, and $\omega = 0.81$. Four cases from rarefied to continuum regimes are simulated, and their computational parameters are shown in Tables 5 and 6, respectively. The Reynolds number is defined as $\text{Re} = \rho U_{lid} L / \mu$, where L is length of the cavity boundary.

For rarefied gas flows (cases 1 and 2), horizontal velocity profiles along AOC (left) and perpendicular velocity profiles along DOB (right) are shown in Fig. 7, respectively. The results of the USP-ESBGK method are consistent with the DSMC results given by ref. [14]. For continuum gas flows (cases 3 and 4), our results obtained by the USP-ESBGK method are consistent with the numerical solutions of NS equation obtained by Ghia [49], as shown in Fig. 8. Note that the time steps in cases 3 and 4 are roughly 3 and 8 times of the mean collision time, respectively. On the other hand, the SP-ESBGK and DSMC methods cannot be used with such large time steps.

**Table 5:** Computational parameters of the square cavity flows (rarefied regime).

| Cases | Kn | $U_{lid}$ (m/s) | $N_{cell}$ | $\Delta t / \tau_c$ |
|---|---|---|---|---|
| 1 | 1.0 | 50 | 56×56 | $0.3\Delta x / \lambda$ |
| 2 | 0.075 | 50 | 56×56 | $0.3\Delta x / \lambda$ |

**Table 6:** Computational parameters of the square cavity flows (continuum regime).

| Cases | Kn | Re | $N_{cell}$ | $\Delta t / \tau_c$ |
|---|---|---|---|---|
| 3 | $1.44 \times 10^{-3}$ | 100 | 72×72 | $0.3\Delta x / \lambda$ |
| 4 | $5.42 \times 10^{-4}$ | 1000 | 72×72 | $0.3\Delta x / \lambda$ |

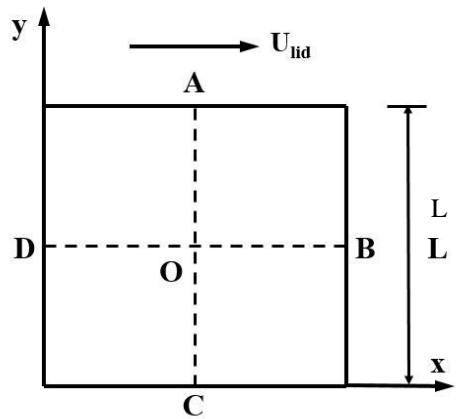

**Figure 6**: Schematic diagram of a square cavity flow.

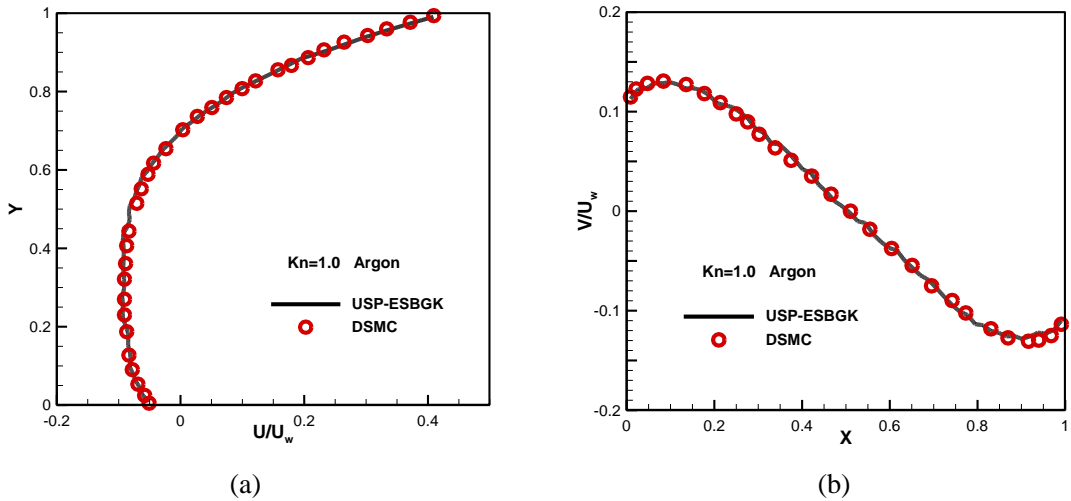

(a)    (b)

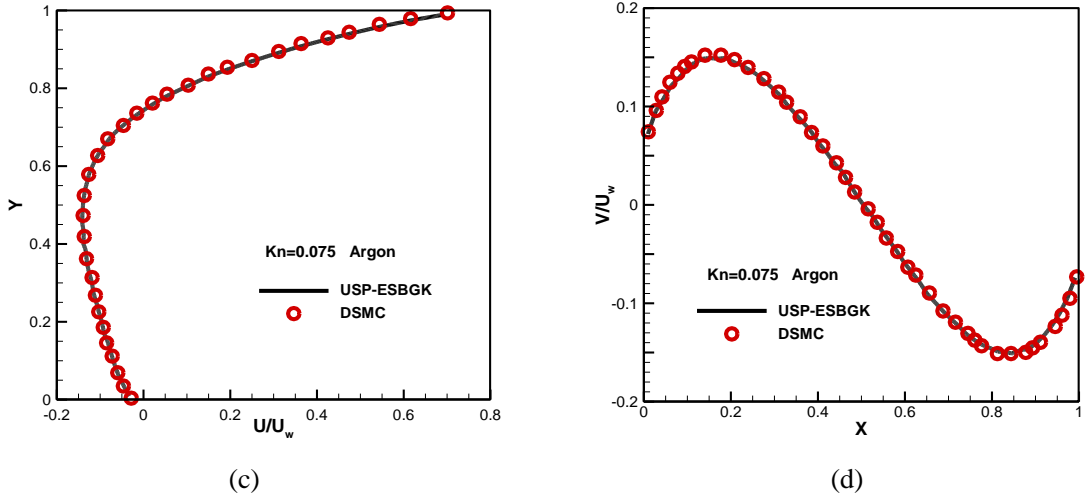

(c)                    (d)

**Figure 7:** (Color online) Horizontal velocity profiles along AOC (left) and perpendicular velocity profiles along DOB (right) and in the square cavity flows for rarefied gas flows at two Kn numbers. Solid line: the present results obtained by the USP-ESBGK method; circle: the DSMC results by Huang et al. [14].

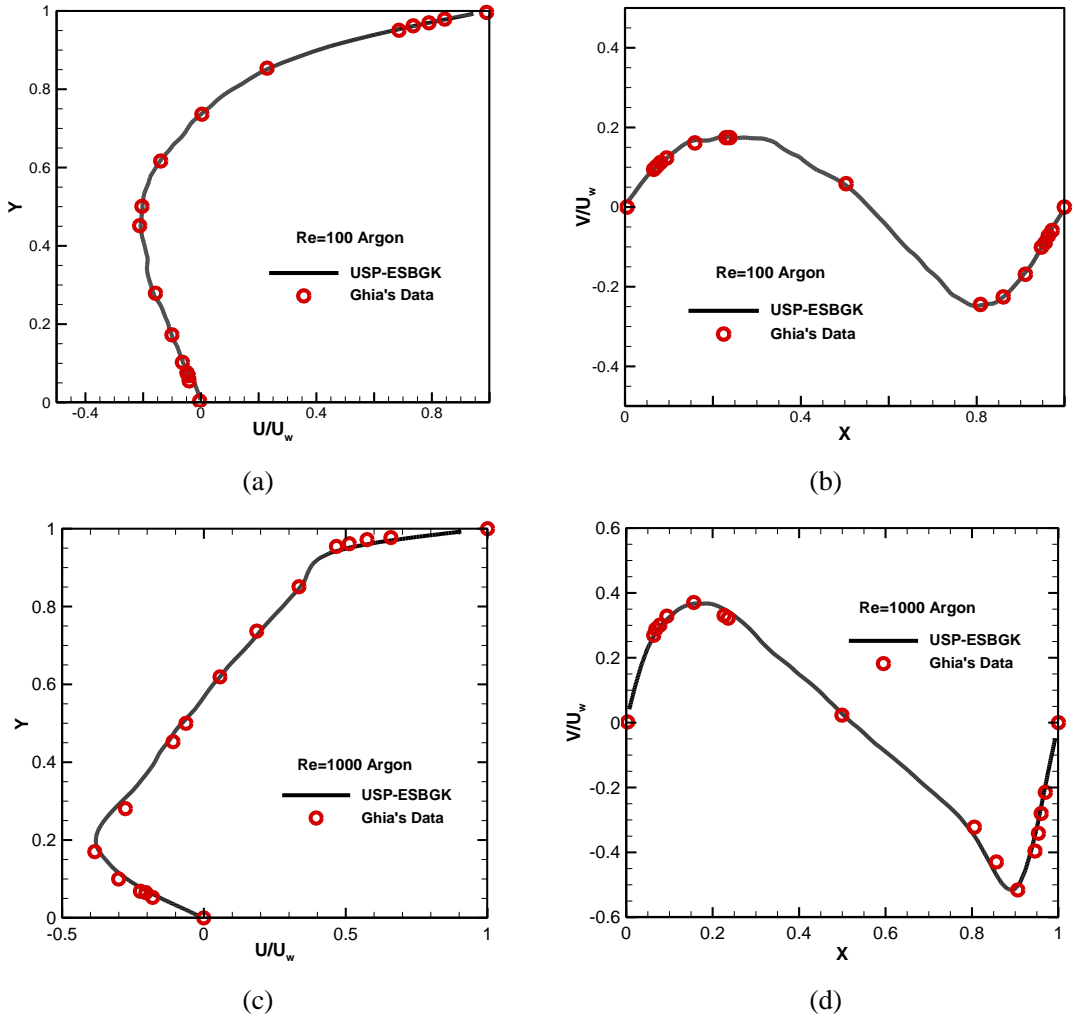

**Figure 8:** (Color online) Horizontal velocity profiles along AOC (left) and perpendicular velocity profiles along DOB (right) in the square cavity flows at two Re numbers (Re=100 and Re=1000). Solid line: results obtained by the USP-ESBGK method; circle: the Navier-Stokes numerical solutions by Ghia [49].

## 4.5 Flow through a slit

Gas flow through a slit has been widely applied from microfluidics to space system. As shown in Figure 9, the computational domain is consisted of two chambers, and the gas flow is generated by the pressure gradient across the slit. The left and right chambers contact with two reservoirs at pressures $p_1 > 0$ and $p_2 = 0$, respectively. The temperature and number density of the left reservoir is $T_1 = 273K$ and $n_1 = 10^{20} m^{-3}$, respectively, and vacuum is assumed for the right reservoir. Therefore, the rarefication of the gas flows increases from left to right chamber. In our simulations, Argon gas with the hard sphere model is used. The computational domain has a height of $L = 20a$, where $a$ is the slit width. Similar to ref. [50], the slit width is chosen as

$$a = \frac{\delta \mu_1 v_1}{p_1}, \qquad v_1 = \left(\frac{2k_B T_1}{m}\right)^{1/2}. \tag{40}$$

Rarefaction parameter $\delta$ refers to the reciprocal of the Kn number, and it equals to 20 in our simulations. Here we make use of the symmetry and only compute the upper half of the flow region. Two chambers are separated by an isothermal wall of 273K, and a fully diffusive boundary condition was assumed. The slit flows are simulated by the SP-ESBGK and USP-ESBGK methods with different cell sizes and time steps. The other computational parameters are given in Table 7.

**Table 7:** Computational parameters of the flows through a slit for SP-ESBGK and USP-ESBGK methods.

| Cases | 1 | 2 | 3 | 4 | 5 | 6 |
|---|---|---|---|---|---|---|
| $N_{cell}$ | 580×280 | 580×280 | 580×280 | 280×140 | 160×80 | 160×80 |
| $\Delta t / \tau_{c,1}$ | 0.2 | 1.0 | 2.0 | 3.0 | 4.0 | 5.0 |

The reduced mass flow rate through the slit is defined as

$$W = \frac{\dot{M}}{\dot{M}_{fm}}, \tag{41}$$

where $\dot{M}$ is the mass flow rate across the slit and $\dot{M}_{fm} = ap_1 / (\sqrt{\pi} v_1)$ is the mass flow rate in the free molecular regime for the planar geometry. The reduced mass flow rate obtained by the SP-ESBGK and USP-ESBGK methods are plotted in Figure 10. It is shown that the reduced mass flow rate predicted by the SP-ESBGK method decrease significantly as time step increases. Since the numerical viscosity of the SP-ESBGK method increases for larger time steps, the rarefaction parameter $\delta$ decreases as shown in Eq. (40). As discussed in ref. [50], the reduced mass flow rate decreases to 1.0 when $\delta \to 0$. Therefore, the reduction of $W$ is expected in the SP-ESBGK method for large time step. On the contrary, as molecular convection and collision effects are coupled in the USP-ESBGK method, their transport coefficients and hence the reduced mass flow rate are hardly influenced by the time steps. In addition, the corresponding reduced mass flow rate predicted by DSMC method is about 1.535 reported in ref. [50]. The relative error between our results and DSMC results is less than 1%, when the USP-ESBGK method is applied with fine computational condition.

For the case 4, the ratio between the time step and the local mean collision time $\tau_{c,loc}$ is shown in Figure 11. It can be seen that this ratio varies from 2.5 to 0.05 near the slit. Even for these coarse cell sizes and time steps, the USP-ESBGK method can still obtain reasonable results. The temperature, macroscopic velocities in both $x$ and $y$ directions, and the Mach number of the case 4 obtained by the USP-ESBGK method are shown in Figure 12. These results agree well with those obtained by the SP-ESBGK method, while the SP-ESBGK method is used with much finer computational cell sizes and time steps as the parameters shown in the case 1 of Table 4.

At the same computational mesh sizes and time steps, the computing time of the USP-ESBGK method is a little bit larger than that of the SP-ESBGK method due to the procedure of adding and deletion particles. However, since the USP-ESBGK method can be used with much larger cell sizes and time steps than the SP-ESBGK method, overall, the USP-ESBGK method is much more efficient for the simulation of continuum gas flows as well as multiscale gas flows.

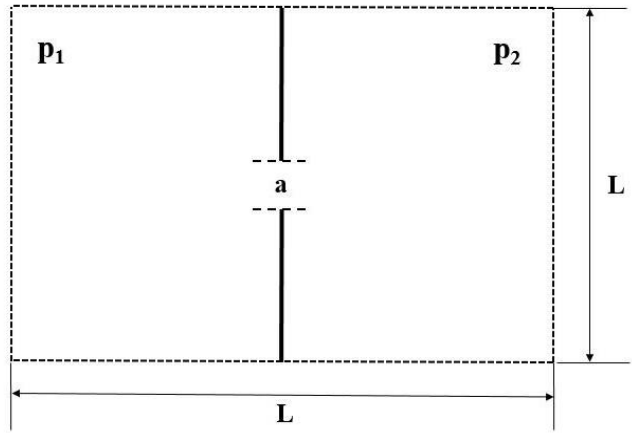

**Figure 9**: Schematic diagram of gas flow through a slit

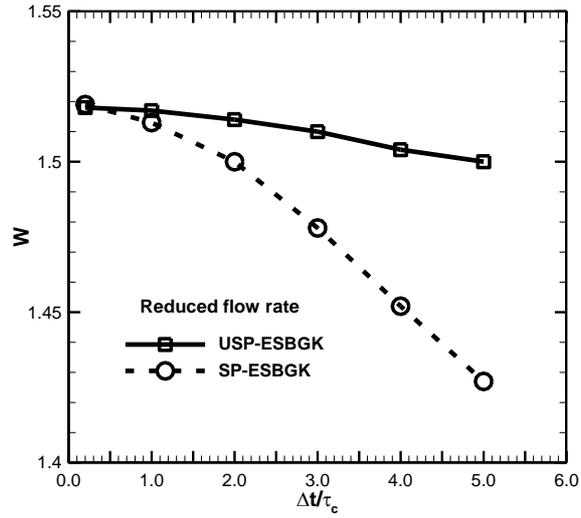

**Figure 10**: Comparison of the reduced mass flow rate of gas flow through a slit between the USP-ESBGK (solid line) and SP-ESBGK (dashed line) methods with different time steps.

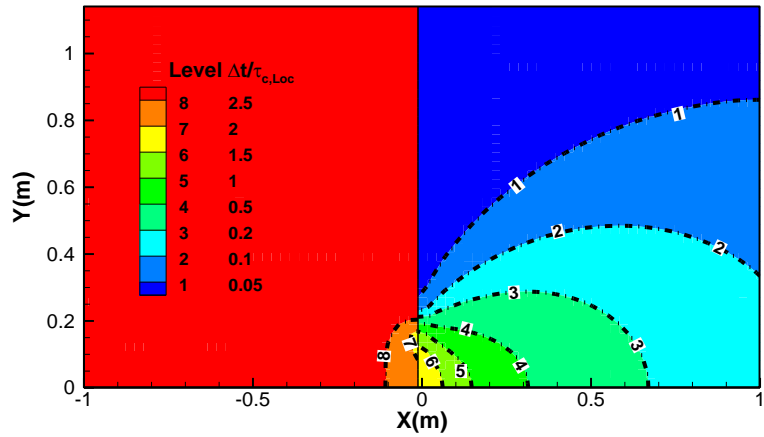

**Figure 11**: (Color online) The ratio of the time step and the local mean collision time for gas flow through a slit obtained by the USP-ESBGK method, for the case 4 in Table 4.

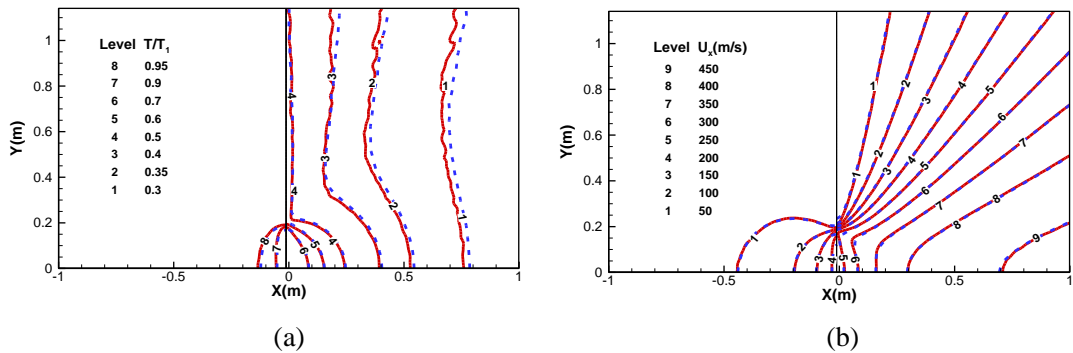

(a)         (b)

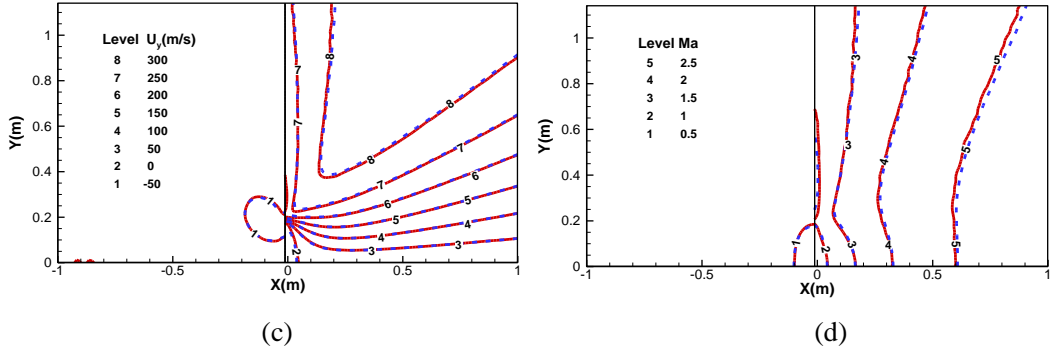

(c)                              (d)

**Figure 12**: (Color online) Gas flow through a slit. (a) Contours of temperature; (b) Contours of streamwise velocity component; (c) Contours of transverse velocity component; (d) Mach contours. The results obtained by the SP-ESBGK method (case 1 in Table 4) and the USP-ESBGK method (case 4 in Table 4) are shown in red solid lines and blue dashed lines, respectively.

## 5. Conclusions

In the present paper, a unified stochastic particle method based on the ESBGK model (USP-ESBGK) has been proposed for the simulation of multiscale gas flows. Several 1-D and 2-D benchmark problems including the Couette flow, thermal Couette flow, Poiseuille flow, and cavity flow have been simulated using the USP-ESBGK method in both rarefied and continuum regimes to check its validity. Furthermore, two typical multi-scale gas flows, i.e. the Sod tube flow and flow through a slit, have also been simulated using the proposed method. Compared with the traditional SP-ESBGK method, the USP-ESBGK method improves the prediction of flow quantities significantly for large temporal-spatial discretization. By combining the molecular convection and collision effects in the simulation, the USP-ESBGK method is able to simulate the small scale non-equilibrium and large scale continuum gas flows in a unified computational framework.

Similar to the current particle/particle (SP-ESBGK and DSMC) hybrid method, it is natural to develop a USP-ESBGK and DSMC hybrid method for the simulation of multiscale gas flows. Since the USP-ESBGK method can be used with much larger temporal-spatial discretization, it is more efficient than the SP-ESBGK method especially in the continuum regime. Therefore, the USP-ESBGK and DSMC hybrid method is a promising tool for the simulation of complex multiscale gas flows. This work will be done in the future.

## Acknowledgements

This work was supported by the National Natural Science Foundation of China (Grants No. 51506063, No. 11772034 and No. 51390494), the Foundation of State Key Laboratory of Coal Combustion (FSKLCCB1702), and the Fundamental Research Funds for the Central

Universities (No. 2014QN183). Results were obtained using National Supercomputer Center in Guangzhou.

## Appendix A: The asymptotic property of the USP-ESBGK collision term at large spatial-temporal scale

For gas flows with large spatial-temporal scales ($Kn_{GLL,MAX} \to 0$), $P_{ne}$ is expanded as

$$P_{ne} = 1 - Kn_{GLL,MAX}/Kn_c + O(Kn^2_{GLL,MAX}). \tag{A1}$$

Hence, the first order of the Chapman-Enskog expansion for the collision term (Eq. (14)) reads

$$J^{(1)}_{(USP-ESBGK)} = -\upsilon f_e \left[ \frac{\sigma^{(1)}_{ik}}{2p} \frac{C_{<i}C_{k>}}{\theta} + \frac{2}{5}\frac{q^{(1)}_k}{p\theta}\Pr C_k \left( \frac{C^2}{2\theta} - \frac{5}{2} \right) \right], \tag{A2}$$

where the stress and heat flux are also expanded as

$$\sigma_{ij} = \varepsilon \sigma^{(1)}_{ij} + \varepsilon^2 \sigma^{(2)}_{ij} + \cdots, \text{ and } q_i = \varepsilon q^{(1)}_i + \varepsilon^2 q^{(2)}_i + \cdots. \tag{A3}$$

The parameter $\varepsilon$ is a formal smallness parameter, which plays the role of the Knudsen number for monitoring the order of terms. Similarly, the collision term of the ESBGK model can also be analyzed by the Chapman-Enskog expansion, and its first order satisfies

$$J^{(1)}_{(ESBGK)} = \upsilon_{ES} \left( f^{(1)}_G - f^{(1)} \right). \tag{A4}$$

Substituting the first order expression for the velocity distribution function of the ESBGK model [44], Eq. (A4) can be rewritten as

$$J^{(1)}_{(ESBGK)} = -\upsilon f_e \left[ \frac{\sigma^{(1)}_{ik}}{2p} \frac{C_{<i}C_{k>}}{\theta} + \frac{2}{5}\frac{q^{(1)}_k}{p\theta}\Pr C_k \left( \frac{C^2}{2\theta} - \frac{5}{2} \right) \right]. \tag{A5}$$

Comparing Eqs. (A2) and (A5), it indicates that the first order Chapman-Enskog expansion of the assumed collision term in the USP-ESBGK method is identical to that of the ESBGK collision term. Therefore, the assumed collision term in the USP-ESBGK method also satisfies the NS solution for large spatial-temporal scales.

## Appendix B: The asymptotic property of the USP-ESBGK collision term at small spatial-temporal scale

For gas flows with small spatial-temporal scales ($Kn_{GLL,MAX} \to +\infty$), the assumption of 13 moments Grad's distribution in the collision term $J_{(USP-ESBGK)}$ is invalid. However, as $P_{ne} \to 0$, $J_{(USP-ESBGK)} \to 0$, and the collision term in Eq. (19b) approaches to Eq. (7b) in the SP-ESBGK method. Therefore, the PDF of the simulated particles turns to be directly solved by the traditional SP-ESBGK method, which has been demonstrated to be accurate enough for these small scale gas flows.